\documentclass[10pt,conference]{IEEEtran}

\usepackage[T1]{fontenc}					% encodage des polices
\usepackage[latin1]{inputenc} 		% 
\usepackage{ae,aecompl}						% use vectorial fonts for adobe acrobat
\usepackage{amsmath}							% package ams
\usepackage{amsfonts}							% polices ams supplémentaires
\usepackage{latexsym}							% symboles supplémentaires
\usepackage{graphicx}
\usepackage{color}
\usepackage{psfrag}
\usepackage{cite}									% Biblio IEEE 
\newcommand{\Iota}{I}
\newcommand{\TD}{TD}
\newcommand{\JD}{JD}

\title{Analysis and design of raptor codes for joint decoding using Information Content evolution}
\author {
					\authorblockN{Auguste Venkiah, Charly Poulliat, David Declercq}  
					\authorblockA{ETIS - CNRS UMR 8051 - ENSEA - University of Cergy-Pontoise}
					{\tt email:\{venkiah,poulliat,declercq\}@ensea.fr} }  

\begin{document}
\maketitle	

\begin{abstract}
In this paper, we present an analytical analysis of the convergence of raptor codes under joint decoding  
over the binary input additive white noise channel (BIAWGNC), 
and derive an optimization method. 
We use Information Content evolution under Gaussian approximation, 
and focus on a new decoding scheme that proves to be more efficient: 
the joint decoding of the two code components of the raptor code.
In our general model, the classical tandem decoding scheme appears to be a subcase, 
and thus, the design of LT codes is also possible. 

\textit{Keywords: Raptor code, joint decoding, optimization of distribution}
\end{abstract}

\section{Introduction}
%###################################################################################################################
\label{secIntroduction}
Fountain codes were originally introduced \cite{Byers} to transmit efficiently over an erasure channel 
with unknown erasure probability.
They are of great interest for multicast or peer to peer applications, and when no feedback channel is available. 

LT codes are the first class of efficient fountain codes, introduced by Luby \cite{LTcodes}.
An LT code produces a potentially limitless number of independent output symbols according to an output degree distribution. 
LT codes are proved to be asymptotically capacity achieving on the Binary Erasure Channel (BEC) \cite{LTcodes}, \cite{raptorPaper}. 
High performance is achieved by designing good output degree distributions.
In order to obtain arbitrary small decoding failure probability, the average degree of the output symbols has to grow
at least logarithmically with $k$, the number of input symbols. 
Thus, performance is achieved at a decoding cost growing in  $O(k \log(k))$. 
This complexity is too high to ensure linear encoding and decoding time which is a desired property for practical codes.

Raptor codes are a class of fountain codes introduced by Shokrollahi in \cite{raptorPaper} as an extention of LT codes.
A raptor code is the concatenation of an LT code and an outer code, called precode.
The precode is a very high rate error correcting block code. 
Thus, the condition of recovering each and every input symbol with arbitrarily high probability can be relaxed: 
the LT code needs to recover a large enough proportion of input symbols, 
and the precode is in charge of recovering the fraction of input symbols unrecovered by the LT code. 
This enables the design of degree distributions of constant mean {\em i.e.} linear encoding and decoding time. 
In \cite{onlineCodes}, the author independently presented the idea of precoding to obtain linear decoding time codes. 
Recently in \cite{Etesami}, 
the results over the BEC of \cite{raptorPaper} were extended to general binary memoryless symmetric channels.  

In all the previously proposed approaches, the LT code and the precode are decoded separately. 
In this paper, we consider another decoding scheme: the joint decoding of the two code components. 
The main idea behind joint decoding is that the precode can help the LT code to converge, by providing extrinsic information. 
By taking into account the information provided by the precode, 
the optimization problem of an LT code becomes less constrained, 
and for a given precode, the total achievable rate of the raptor code becomes closer to the channel capacity.

In this paper, we develop the asymptotic analysis of the joint decoder, and propose an optimization method for the design of efficient 
degree distributions. 
For this purpose, we use a fully analytical approach: information content (IC) evolution under Gaussian approximation (GA).
We introduce the extrinsic transfer function of the precode into the equations, 
which leads to a new model that takes into account the information provided by the precode. 
In our analysis, the classical separate decoder appears to be a sub-case of the joint decoder, 
by assuming that no extrinsic information is passed from the precode to the LT code. 

The remainder of this paper is organized as follows: 
In section \ref{secSystemDescription}, we descibe the system that we consider and give the notations used in the paper.
In section \ref{secAsymptoticAnalysisDesign}, we study the asymptotic performance of raptor codes on the BIAWGNC, 
state the optimization problem for the design of output degree distributions, and analyze the main design parameters.  
In section \ref{secResults}, we show experimental results.

%###################################################################################################################
\section{System description and notations}
\label{secSystemDescription}
%------------------------------------------------------------------------------------------------------------------
\subsection{LT codes and raptor codes}
We call {\em input symbols} the set of information symbols to be transmitted 
 and {\em output symbols} the symbols produced by an LT code from the input symbols. 
An LT code is described by its {\em output degree distribution}. 
To generate an output symbol, a degree $d$ is sampled from that distribution, independently from the past samples. 
The output symbol is then formed as the binary sum of a uniformly randomly chosen subset of size $d$ of the input symbols: 
the $d$ input symbols and the output symbols verify a parity check equation. 

Let $\Omega _1, \Omega _2 , \dots , \Omega _{d_c}$ be the distribution weights on $1,2, \dots , d_c$ so that $\Omega _d$ denotes 
the probability of choosing the value $d$ under this distribution. 
We denote the output degree distribution using its  generator polynomial:  
$\Omega (x) = \sum _{i=1}^{d_c} \Omega _i x^i$, 
which is associated with the corresponding edge degree distribution 
$\omega(x) = \sum _{i=1}^{d_c} \omega _i x^{i-1} = \Omega '(x)/\Omega '(1)$. 

Because input symbols are chosen uniformly at random, their node degree distribution is binomial, 
and can be approximated by a Poisson distribution with parameter $\alpha$ \cite{raptorPaper}. 
Thus, we have the polynomial that describes the input symbols degree distribution defined as: 
$$ \Iota (x) = e^{\alpha (x-1)} $$
Moreover, the associated input edge degree distribution 
$\iota (x) = \sum _{i=1}^{d_v} \iota _i x^{i-1} = \Iota '(x)/ \Iota '(1)$ also equals $e^{\alpha (x-1)}$. 
Both distributions are of mean $\alpha$. 

Input symbols are not transmitted over the channel. 
At the receiver side, we have noisy observations of the output symbols, 
and belief propagation (BP) decoding is used to recover iteratively the input symbols.

A raptor code is an LT code concatenated with an outer code called ``precode''. 
The input symbols of the LT code are then formed by a codeword of the precode. 

Although fountain codes are rateless, we can define the {\em a posteriori} rate R of an LT code as follows:
\begin{eqnarray}
				R_{LT}  \nonumber &=& \frac{\text{\small Nb input symbols}}
				{\text{\small Nb output symbols needed for successfull decoding}} \nonumber \\
				&=& \frac{\Omega '(1)}{\alpha}
				\label{eqDefR}
\end{eqnarray}

\begin{figure}[t]
				\centering
				\includegraphics[width=0.48\textwidth]{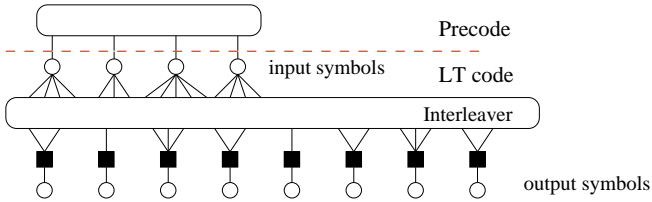} 
				\caption{Description of a raptor code:  Tanner graph of an LT code + precode.
				The black squares represent the parity check nodes and the circles represent variable nodes
				associated with input symbols or output symbols.}
				\label{figLTtannerGraph}
\end{figure}

As for LDPC codes, a raptor code can be represented by a Tanner graph. 
A Tanner graph is a bipartite representation of a system composed of data nodes and function nodes. 
Here, the data nodes represent input or output symbols 
and the function nodes represent how their adjacent data nodes interact through parity checks.
The edges on the graph carry probability messages that come in or out of the data nodes. 
The Tanner graph of a raptor code is given in Fig. \ref{figLTtannerGraph}. 

\subsection{Tandem and joint decoding of a raptor code}
The classical``Tandem decoding'' (\TD) consists of decoding the LT code first and then using the extrinsic information about the input symbols 
as {\em a priori} information for the precode. 

For ``Joint decoding'' (\JD), one decoding iteration consists of alternating 
$N_{lt}$ decoding iterations on the LT code, 
and $N_p$ decoding iterations on the precode. 
Thus, both code components of the raptor code provide extrinsic information to each other.
In the sequel, we shall only consider the case where $N_{lt} = N_p = 1$, 
and where the precode is an LDPC code. 
In this particular case, %it can be shown that 
the raptor code can be described by a single Tanner graph 
with two kinds of parity check nodes : 
check nodes of the precode, referred to as ``static check nodes'' and  
parity check nodes of the LT code, later referred to as ``dynamic check nodes''.

Because the precode provides extrinsic information to the LT code, 
we need to introduce the extrinsic transfer function of the precode, 
denoted by $x \mapsto T(x)$, into the IC evolution equations.

\section{Asymptotic analysis and design of raptor codes for the BIAWGNC} \label{secAsymptoticAnalysisDesign}
%###################################################################################################################
In this section, we derive the asymptotic analysis of a raptor under \JD. 
Thus we assume that extrinsic information is exchanged between the precode and the fountain at each decoding iteration. 
The analysis will be presented from the fountain point of view, 
and we will track the evolution of the IC of the messages that are related to the fountain part of the Tanner graph. 
Indeed, our objective is to optimize the distribution of the fountain part of the raptor code, 
namely $\omega(x)$, taking into account the contribution of the precode through its IC transfer function.

For our study, we use IC evolution under GA and treelike assumption. 
This allows us to keep a fully analytical and monodimensionnal approach, 
without the need for Monte Carlo simulations as done in \cite{Etesami}, 
thus leading to a more computationnally efficient optimization. 
IC evolution is a concurrent tool of mean evolution under GA \cite{chung}, 
that has been proved to be more accurate and robust for the optimization of LDPC/IRA codes \cite{ira}. 

The messages on the decoding graph are the log density ratios (LDR) of the probability weights. 
They are modeled by a random variable which is assumed to be Gaussian distributed 
with mean $m$ and variance $\sigma ^2=2m$ \cite{chung}. 
Thus, the density of the messages is symmetric \cite{RU}. 
For a message sampled from such a symmetric Gaussian distribution, 
the IC associated to the message is $x = J(m)$ \cite{ira}, where $J(.)$ is defined by:
\begin{equation}
				J(m) = 1- \frac{1}{\sqrt{4\pi m}} \int _{R} \log _2(1+e^{-\nu})\exp\bigg(-\frac{(\nu -m)^2}{4m}\bigg) d\nu
				\label{eqJ}
\end{equation}

\subsection{Asymptotic analysis of raptor codes} \label{subsecEXIT}
%------------------------------------------------------------------------------------------------------------------
\subsubsection{Information content evolution} 
When the precode is an LDPC code with node and check edge distributions $\lambda (x)$ and $\rho (x)$, 
its IC transfer function \cite{TB} is given by: 
\begin{equation}
				T(x) = \sum_{i=2}^{d_c} \lambda_i J \bigg(iJ^{-1} \Big (1-\sum_{j=2}^{d_v}\rho_j J\big ((j-1)J^{-1}(1 - x)\big )\Big )\bigg)
				\label{EXITldpc}
\end{equation}

We denote $x_u^{(l)}$ (resp. $x_v^{(l)}$ ) the IC associated to messages on an edge connecting 
a dynamic check node to an input symbol (resp. an input symbol to a dynamic check node) at the $l^{\text{th}}$ decoding iteration. 
We denote by $x_{\text{ext}}^{(l)}$ the extrinsic information passed by the LT code to the precode, 
at the $l^{\text{th}}$ decoding iteration. 
As the input symbols are of average degree $\alpha$, we have: 
\begin{equation}
				\nonumber x_{ext}^{(l)}=J\big(\alpha J^{-1}(x_u^{(l)})\big)
\end{equation}
The extrinsic information passed by the precode to the LT code is then $T(x_{\text{ext}}^{(l)})$. 
When accounting for the transfer function of the precode, 
the IC update rules for the IC in the Tanner graph can be written as follows: 
\begin{align}
				x_v^{(l)}&= \sum_{i=1}^{d_c}\iota_{i}J\Big((i-1)J^{-1}(x_u^{(l-1)})+J^{-1}\big(T(x_{ext}^{(l-1)})\big)\Big)\label{eqXvu} \\
				1-x_u^{(l)}&= \sum_{j=1}^{d_v}\omega _j J \Big ((j-1)J^{-1}(1 - x_v^{(l)}) + f_0 \Big ) \label{eqXv2Xu} 
\end{align}
with:
$$f_0 = J^{-1}\Big ( 1- J\Big(\frac{2}{\sigma ^2}\Big)\Big )$$

\begin{figure*}[t]
\begin{equation}
				x_u^{(l)} = F(x_u^{(l-1)}, \sigma ^2) = 1 - \sum_{j=1}^{d_v}\omega _j J \bigg ((j-1)J^{-1}\Big(1 - \sum _{i=1}^{d_c} \iota _i J \big ((i-1)J^{-1}(x_u^{(l-1)}) +J^{-1}\big(T(x_{ext}^{(l-1)})\big) \big ) \Big ) + f_0 \bigg ) \label{eqF} 
\end{equation}
\hrulefill
\end{figure*}
Replacing \eqref{eqXvu} in \eqref{eqXv2Xu} gives   \eqref{eqF}, 
that describes the evolution through one joint decoding iteration of the IC of the LDRs at the output of the dynamic checknodes (fountain part):  $x_u^{(l)} = F(x_u^{(l-1)}, \sigma ^2)$.
Note that for a given distribution $\iota (x)$, this expression is linear with respect to the coefficients of $\omega (x)$, 
which is the distribution that we intend to optimize. 

We point out that \eqref{eqF} is general since it reduces to the classical \TD\ case 
by setting the extrinsic transfer function to $x \mapsto T(x)=0 \quad \forall x \in [0;1]$, 
thus assuming that no information is propagated from the precode to the fountain.

\subsubsection{Fixed point caracterization}
\label{subsecFixedPoint}
In an IC evolution analysis, the convergence is guaranteed by $ F(x, \sigma ^2)> x$. 
Convergence continues toward a fixed point of $x \mapsto F(x, \sigma ^2)$. 
Unfortunately, there are no trivial solutions for the fixed point of \eqref{eqF}. 
However, using a functionnal analysis, an upper bound on the fixed point can be given. 
Replacing $x_u$ by $1$ and using the fact that $T(1)=1$ in \eqref{eqF}, we obtain: 

\begin{equation}
				\lim _{x \rightarrow 1} F(x, \sigma ^2) = J\Big(\frac{2}{\sigma ^2}\Big) = x_0
				\label{eqXo}
\end{equation}
which means that, because $x \mapsto F(x, \sigma ^2)$ is an increasing function, 
the fixed point is necessarily less or equal than $x_0$, 
which is the capacity of a BIAWGNC with parameter $\sigma ^2$.  
Thus, the IC is upper bounded through the decoding iterations by $x_0$. 
This gives some insights on the asymptotic behavior at the decoding convergence point: 
the BP decoding of the LT part of a raptor code is limited on a BIAWGNC by the capacity of the channel. 

This result is not really surprising and can be interpreted as follows: 
the output nodes of degree one have a constant contribution on each check node.  
As the iterative decoding process goes on, the IC of the messages 
at the output of the dynamic check nodes is limited by the channel observations.

\subsubsection{Starting condition}
We now derive a condition for the beginning of the decoding process: at the first iteration, $x_u^{(0)}=0$. 
Therefore, according to \eqref{eqXvu}, $x_v^{(1)}=0$.
Reporting this in \eqref{eqF} gives:
\begin{equation}
				x_u^{(1)} = F(0,\sigma ^2) = \omega _1 J\bigg(\frac{2}{\sigma ^2}\bigg) \nonumber
				\label{eqF0}
\end{equation}
The decoding process can begin iff $x_u^{(1)}>\varepsilon $, for some arbitrary $\varepsilon > 0$, 
which gives: 
%Then, the starting condition constraint on $\omega _1$ is given by: 
\begin{equation}
				\omega _1 > \frac {\varepsilon }{ J\big(\frac{2}{\sigma ^2}\big)}
				\label{eqConditionW1}
\end{equation}

Therefore, one must have $\omega _1 > 0$ for the decoding process to begin, 
and $\varepsilon$ appears to be a design parameter that will constrain the optimization problem, 
ensuring that $\omega_1 \ne 0$. 
In practice, the value of $\varepsilon$ can be chosen arbitrarily small. 
Indeed, it has been proved \cite{Etesami} that for a sequence of capacity achieving distributions $\omega^{(n)} (x)$, 
$\lim_{n \rightarrow \infty}\omega^{(n)}_1 = 0$.

Remark: as an illustration we point out that, for the ``Ideal Soliton Distribution'' introduced by Luby \cite{LTcodes}, 
$\Omega _1 = 1/k$, which is the smallest proportion possible with $k$ input symbols. 

\subsubsection{Lower bound on the edge proportion of degree 2 output symbols}
For an output degree distribution that is to be capacity achieving, we have: 
\begin{equation}
\omega _2 > \frac{1}{(\alpha-1)e^{-f_0 /4} }
\end{equation}

We only give a sketch of the proof. 
Let $\mu$ be defined by $\mu = J^{-1}(1-x)$. 
By derivating $x \mapsto F(x,\sigma ^2)$ defined in \eqref{eqF}, 
and using the approximation of the derivative of $J(\mu)$ for large $\mu$ given in \cite{ira}, we get:
\begin{equation}
				\lim _{x \rightarrow 0} F'(x,\sigma ^2) = \omega _2 (\alpha -1)e^{-f_0/4} \nonumber
				\label{eqDegres2}
\end{equation}
Moreover, for a capacity achieving degree distribution, $\omega _1 = 0$ \cite{Etesami}, which means that $F(0,\sigma ^2) = 0$.
Then, the convergence condition $F(x,\sigma ^2) > x $ implies that $\lim _{x\rightarrow 0}F'(x,\sigma ^2) > 1$, 
which gives the result.  

Remark: the IC evolution method leads to a slightly different result than the one obtained with mean evolution \cite{Etesami}. 
The same phenomenon has been observed for the derivation of the stability condition, for the optimization of LDPC codes. 

\subsection{Design of output degree distributions} \label{subsecDesignLT}
In this section, we explicit the optimization problem for the design of output degree distributions, 
and give some complementary results that we used for the choice of the design parameters. 
%------------------------------------------------------------------------------------------------------------------
\subsubsection{Optimization problem statement} \label{subsecMatching}
The optimization of an output distribution consists of maximizing the rate of the corresponding LT code, 
{\em i.e.} maximizing $\Omega '(1) = \sum _i \Omega _i i$, which is equivalent to minimizing $\sum _i \omega _i/i$.  
Moreover, according to the previous section, several constraints must be satisfied. 
As $\omega (x)$ is a probability distribution, its coefficients must sum up to 1. 
We call this the proportion constraint $[\mathrm{C}_1]$.   
Moreover, the convergence implies that $F(x, \sigma ^2) > x$.  
However, this inequality cannot hold for each and every value of $x$: 
the analysis in section \ref{subsecFixedPoint} shows that the fixed point of $F(x, \sigma ^2)$ 
is smaller than $x_0 = J\big(\frac{2}{\sigma ^2}\big)$.
Therefore, we must fix a margin $\delta >0$ away from $x_0$.  
By discretizing $[0 ; x_0-\delta]$ and requiring inequality to hold on the discretization points, 
we obtain a set of inequalities that need to be satisfied: they define the convergence constraint $[\mathrm{C}_2]$. 
The starting condition \eqref{eqConditionW1} must also be satisfied and defines the constraint $[\mathrm{C}_3]$.
Moreover, the edge proportion of output symbols must fullfill \eqref{eqDegres2}, 
defining the stability constraint $[\mathrm{C}_4]$.  
Finally, $x \mapsto T(x)$ is defined according to \eqref{EXITldpc} for an LDPC code,
or could be estimated with Monte Carlo simulations if another component code is used as a precode. 

For a given value of $\alpha$, 
the cost function and the constraints are linear with respect to the unknown coefficients $\omega _i$. 
Therefore, the optimization of an output degree distribution can be written as a linear optimization problem 
that can be solved with linear programming.
For a given $\alpha$, the optimization problem can be stated as follows: 
\begin{eqnarray}
				\label{eqOptimization} 
				\omega_{opt}(x)&=&\textrm{arg}\min_{\omega(x)}{\sum_{j}{\frac{\omega_j}{j}}}
\end{eqnarray}
subject to the constraints: 
\begin{itemize}
				\item[$\mbox{[}\mathrm{C}_1\mbox{]}$]$\sum _i \omega _i =1$ 
				\item[$\mbox{[}\mathrm{C}_2\mbox{]}$]$F(x,\sigma ^2)>x\quad\forall x \in [0;x_0 - \delta ]\quad\text{ for some } \delta >0$
				\item[$\mbox{[}\mathrm{C}_3\mbox{]}$]$F(0,\sigma ^2) > \varepsilon \quad \text{ for some } \varepsilon > 0$
				\item[$\mbox{[}\mathrm{C}_4\mbox{]}$]$F'(0,\sigma ^2) > 1 $ 
\end{itemize}

\subsubsection{Lower bound on $\alpha$}
\label{subsubsecBoundAlpha}
The average degree of input symbols $\alpha$ is the main design parameter. 
At the output of an LT code, the IC of messages sent from the LT code to the precode is given by:  
\begin{equation}
				x_{\text{ext}} = J\big( \alpha J^{-1}(x_u^{(\infty)})\big)
				\label{eqXext}
\end{equation}
Moreover, let $x_{\text{p}}$ be the IC threshold above which the decoding of the precode is successfull. 
Then successfull decoding of the raptor code is obtained if $x_{\text{ext}} > x_{\text{p}}$. 
Using equation \eqref{eqXext}, and recalling that $x_u^{(\infty)}< x_0$ we get a lower bound on $\alpha$:
\begin{equation}
				\alpha \ge \frac{\sigma ^2 J^{-1}(x_{\text{p}})}{2} = \alpha_{min}
				\label{eqAlphaMin}
\end{equation}

This bound can be used to limit the search space on $\alpha$. 
Indeed, for increasing values of $\alpha$, we optimize output degree distributions as explained in the previous section. 
It appears that there is a value for $\alpha$ that maximizes the corresponding rate of the LT code.

\subsubsection{Parameter $\delta$}
The convergence of the LT code should be such that at some point of the decoding process, 
$x_{\text{ext}}$ becomes larger than the precode's threshold $x_{\text{p}}$. 
For a given value of $\alpha \ge \alpha _{min}$, 
$\delta$ is such that $J\big(\alpha J^{-1}(x_0 - \delta)\big) \ge x_{\text{p}}$
{\em i.e.} 
\begin{equation}
				\delta \le x_0 - J\bigg(\frac{\sigma^2 J^{-1}(x_{\text{p}})}{2}\bigg )
				\label{eqDeltaMax}
\end{equation}

\subsection{Considerations on the choice of a precode}
In this section, we discuss some important points concerning the choice of the precode, 
which give another justification to why \JD\ should be preferred over \TD\ 
in the perspective of designing efficient raptor codes. 
Let $R_t$ be the rate of the raptor code which is the concatenation of an LT code of rate $R_{LT}$, 
and a precode of rate $R_{p}$. 
We have:
\begin{equation}
 R_t = R_{p} R_{LT} =  R_{p}\frac{\Omega '(1) }{\alpha }
 \label{eqRateRaptor}
\end{equation}

For a channel with capacity $C$, we have  $R_{LT} < C$ for LT codes optimized for the \TD\ scheme.
Thus, $R_p$ appears to be a burden in terms of the total rate of the raptor code. 
Fortunately, the optimization problem becomes less constrained in a \JD\ scheme, 
because the precode provides extrinsic information to the LT code, 
and the optimization for \JD\ leads to $R_{LT} > C$, 
allowing the use of lower rate precodes than in the \TD\ scheme. 

The use of lower rate precodes can be motivated by the fact that the design of very high rate LDPC codes is a difficult problem. 
Even though the optimization of irregularity profiles can give codes with good thresholds, 
the actual design of such codes remains difficult, because their underlying graph is highly connected. 
The higher the rate, the more difficult it is to design a graph with ``few enough'' short cycles.

In the context of \JD, our optimization procedure addresses naturally the problem of the overall rate distribution 
and its repartition between the fountain code and the precode.

\section{Experimental results} \label{secResults} 
We define the overhead of a fountain code as $\epsilon = \big ( \frac{R_t}{C} \big )^{-1} - 1$. 
Thus, an overhead of 0 means that capacity is achieved. 
An overhead of 0.1 means that the rate of the raptor code is 10\% away from the capacity. 
In our simulations, the performance of a raptor code is evaluated by Bit Error Rate (BER) versus overhead. 

\subsection{LT codes}
First, we use our model to design LT codes. 
We recall that this is possible by defining the extrinsic transfer as a null function.

We design an output degree distribution $\Omega_A(x)$, with parameters $\delta = 0.04$ and $\alpha = 21$. 
The optimization was made for a BIAWGNC of capacity $C=0.5$, in order to compare ourselves 
to the distribution proposed in \cite[p 2044]{Etesami}, referred to as $\Omega _E (x)$. 
As we only test the LT code, we did not use any precode. 
The simulations  were set to $k=65000$ input symbols, and 300 decoding iterations, on a channel of capacity $C=0.5$ ($\sigma = 0.9787$). 

\begin{figure}[h]
	\begin{center}
					\psfrag {Omega}{ { \tiny {$\Omega_A$}} \ \ }
					\psfrag {OmegaE}{ { \tiny {$\Omega_E$}} \ \ }
		\includegraphics[width=0.48\textwidth]{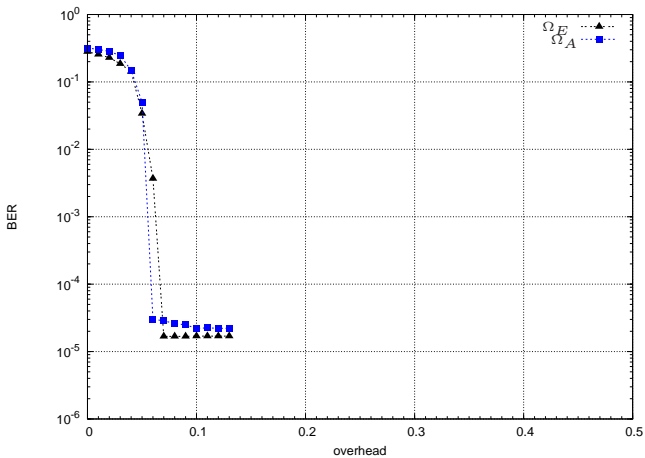}
	\end{center}
	\caption{BER versus overhead for LT codes defined by the distribution $\Omega_A(x)$ optimized for a BIAWGNC of capacity $C=0.5$.
	We compare our distribution to the one proposed in \cite[p 2044]{Etesami}, denoted by $\Omega_E(x)$, with $k=65000$ input bits}
	\label{fig1}
\end{figure}

In Fig \ref{fig1}, we report the BER versus overhead for LT codes defined by $\Omega _A (x)$ and $\Omega _E(x)$. 
Our method appears to be as efficient as the one proposed in \cite{Etesami}, 
but it is computationnally more efficient, since it does not require Monte Carlo simulations. 

\subsection{\TD\ versus \JD}
We now compare \JD\ and \TD\ schemes. 
We used a regular (3,60) LDPC precode of length $N=65000$, generated randomly. 
We compare the distribution $\Omega_E (x)$ proposed in \cite[p 2044]{Etesami} in both \TD\ and \JD\ decoding schemes, 
to a distribution $\Omega_B (x)$ that we optimized for \JD\ with our method. 
For the distribution $\Omega _E (x)$ there is very little difference between \TD\ and \JD\ decoding schemes. 
This can be explained by the fact that the distribution has not been optimized to take into account the information provided 
by the precode. 
For our distribution $\Omega _B (x)$, performance is improved. 
The effect of the precode is to help the convergence of the LT code, which we can interpreted as follows: 
the BER decreases with a slope, whereas for $\Omega _E (x)$, there is clearly  a threshold behavior. 

\begin{figure}[h]
	\begin{center}
					\psfrag {etesamiS}{  {\tiny {$\Omega_E$\ \ \ }}}
					\psfrag {etesamiC}{  {\tiny {$\Omega_E$}}\ \ \ }
					\psfrag {conjR360}{  {\tiny {$\Omega_B$ }}\ \ \ }
		\includegraphics[width=0.48\textwidth]{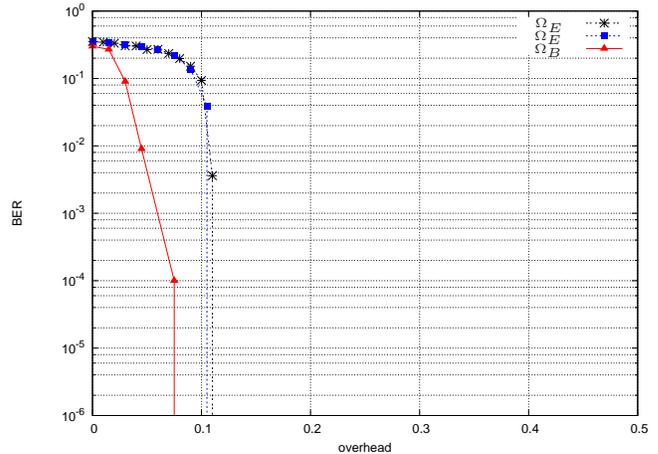}
	\end{center}
	\caption{BER versus overhead for a raptor code defined with a regular (3,60) LDPC precode. 
	We compare $\Omega _B (x)$, a distribution that we optimized for joint decoding, 
	to $\Omega _E(x)$ proposed in \cite{Etesami} under TD (blue squares) and under JD (black stars)}
	\label{figAlpha}
\end{figure}

\section{Conclusion} 
\label{secConclusion}
We presented the analytical analysis of raptor codes with IC evolution under GA, 
stated the optimization problem for the design of output degree distributions well adapted to joint decoding,
and analyzed the main design parameters. 
Our model also allows to design efficient LT codes. 
Experimental results show that \JD\ is more efficient than the classical \TD\ decoding scheme. 

%###################################################################################################################
% Bibliographie:
%###################################################################################################################
\bibliographystyle{IEEEbib} % 

\end{document}